\documentclass[preprint,aps,draft]{revtex4}

\usepackage{graphicx}
\usepackage{dcolumn}
\usepackage{bm}

\begin{document}

\title{Coherent-Squeezed State Representation
of Travelling General Gaussian Wave Packets}

\author{Sang Pyo Kim}\email{sangkim@kunsan.ac.kr}

\affiliation{Department of Physics, Kunsan National University,
Kunsan 573-701, Korea}

\affiliation{Asia Pacific Center for Theoretical Physics, Pohang
790-784, Korea}

\date{\today}
\begin{abstract}
Using the time-dependent annihilation and creation operators, the
invariant operators, for a free mass and an oscillator, we find
the coherent-squeezed state representation of a travelling general
Gaussian wave packet with initial expectation values, $x_0$ and
$p_0$, of the position and momentum and variances, $\Delta x_0$
and $\Delta p_0$. The initial general Gaussian wave packet takes,
up to a normalization factor, the form $e^{i p_0 x/\hbar} e^{- (1
\mp i \delta) (x - x_0)^2 / 4 (\Delta x_0)^2}$, where $\delta =
\sqrt{ (2\Delta x_0 \Delta p_0/\hbar)^2 -1}$ denotes a measure of
deviation from the minimum uncertainty or the initial
position-momentum correlation $\delta = 2\Delta (xp)_0 / \hbar$.
The travelling Gaussian wave packet takes, up to a time-dependent
phase and normalization factor, the form $e^{i p_c x/\hbar} e^{-
(1 - 2 i \Delta (xp)_t/\hbar) (x - x_c)^2 / 4 (\Delta x_t)^2}$ and
the centroid follows the the classical trajectory with $x_c(t)$
and $p_c(t)$. The position variance is found to have additionally
a linearly time-dependent term proportional to $\delta$ with both
positive and negative signs.
\end{abstract}

\maketitle

\section{Introduction}

Localized wave packets have been used as an important method to
describe quantum motions of matters since the advent of quantum
theory (for review and references, see
\cite{buchleitner,robinett}). Recently wave packets have attracted
much attention, in particular, in quantum wave packet revivals
predicted for long time evolution of bounded state systems
\cite{parker} and their experimental confirmation \cite{yeazell}.
The collapse and revival of matter field has also been observed in
a Bose-Einstein condenstate \cite{greiner}. The simplest and most
well-known wave packet is the Gaussian wave packet which is useful
not only to understand concepts of wave packets but also to study
exactly  a certain type of quantum motions.  In the gravitational
wave detection, for instance, a measuring apparatus is a free mass
or an oscillator and the system should be monitored up to quantum
limit due to extreme weak gravitational waves
\cite{braginsky,thorne,braginsky2}. There the quantum motion of
the free mass or oscillator that will provide important
information about the external forces can be described by wave
packets.

The Gaussian wave packet of minimum uncertainty is particularly of
interest from the measurement point of view. It has been known
that the Gaussian wave packet can have arbitrary position and
momentum expectation values and the position variance or momentum
variance \cite{bohm,gottfried,schiff,messiah,cohen}. Thus there
seem to be at most three parameters to fix the profile of the
Gaussian wave packet. However, the variances of the position and
momentum can take arbitrary values as far as the uncertainty
principle is satisfied. As Bohm has already explained \cite{bohm},
the independent variances of the position and momentum necessarily
mean a correlation between these two noncommuting operators. This
conditional freedom in choosing the variances of the position and
momentum may raise the question of which wave packet will have all
four parameters, $x_0$, $p_0$, $\Delta x_0$ and $\Delta p_0$ under
the provisions of $\Delta x_0 \Delta p_0 \geq \hbar/2$. Howard and
Roy made an attempt to extend the Gaussian wave packet by
introducing a position squared phase \cite{howard}. This
additional position-dependent phase factor results in the
position-momentum correlation and is recognized as a squeezing
effect \cite{ford,robinett2}.

On the other hand, various methods to construct quantum states and
techniques to measure them have been developed in quantum optics
and atomic physics (for review and references, see \cite{mandel}).
Glauber's coherent state is a useful tool to find quantum states,
particularly, of an oscillator that displace the initial wave
packets and thus exhibit a classical feature \cite{glauber}. The
squeezed states deform the shape of wave packets by changing one
variance at the price of the other variance \cite{stoler}. There
is one complex parameter, equivalent to two real parameters, for a
coherent state, and one squeeze parameter and one squeeze angle
for a squeeze state. Then a relevant question is whether there is
any connection between four parameters for initial wave packets
and four parameters for the coherent and squeezed state. Years ago
Yuen introduced such a coherent-squeezed state at the initial time
to account for a negative contribution to the position variance
from the initial correlation between the position and momentum
\cite{yuen}. However, there still remains unfound the exact form
of the travelling general Gaussian wave packet evolved from an
initial wave packet.

To exploit the connection between the parameters for Gaussian wave
packets and the parameters for coherent and squeezed states, we
shall employ the invariant operator method introduced by Lewis and
Riesenfeld \cite{lewis}. Though this method was introduced for
time-dependent quantum systems, in particular, oscillators with
time-dependent mass and/or frequency, it can be applied even to
time-independent systems. The idea is that any operator satisfying
the quantum Liouville-von Neumann equation provides exact
solutions of the time-dependent Schr\"{o}dinger equation up to
time-dependent phase factors. For time-dependent oscillators, both
quadratic invariant operators \cite{lewis,hartley,ji1,pedrosa} and
linear invariant operators
\cite{dod,kim1,kim-lee,nieto,kim-page,kim2,kim3} were used to find
the wave functions.

The purpose of this paper is to apply the invariant operator
method to a free mass and an oscillator and find the travelling
general Gaussian wave packet with four parameters. For that
purpose we find a pair of linear invariant operators, the
time-dependent annihilation and creation operators, for the free
mass and the oscillator and then use them to find a
coherent-squeezed state that is an exact solution of the
time-dependent Schr\"{o}dinger equation. And then we relate one
complex parameter for the coherent state and two real parameters
for the squeezed state with the four parameters for the initial
Gaussian wave packet. This identification provides the explicit
form of travelling general Gaussian wave packet that has the
required initial parameters.

The organization of this paper is as follows. In Sec. II, it is
shown that there are at most four parameters for initial Gaussian
wave packets. In Sec. III, we apply the invariant operator method
to find the time-dependent annihilation and creation operators for
a free mass and an oscillator.  It is shown that a general
annihilation operator is a unitary transform of some preferred one
via a squeeze operator. In Sec. IV, the general wave packet is
obtained as a coherent-squeezed state of some preferred Gaussian
wave packet. In. Sec. V, we determine one complex parameter, one
squeeze parameter and angle in terms of four parameters for an
initial Gaussian wave packet. And then we find the evolution of
variances of the position and momentum.

\section{Gaussian Wave Packets}

In the Schr\"{o}dinger picture, many textbooks (for instance, see
\cite{schiff,messiah,cohen}) use the Gaussian wave packet in the
position space
\begin{equation}
\psi(x, t=0) = \frac{1}{(2 (\Delta x_0)^2)^{1/4}} e^{ip_0x/\hbar}
e^{- (x - x_0)^2/4 (\Delta x_0)^2}, \label{ga wa}
\end{equation}
and in the momentum space
\begin{equation}
\phi(p, t=0) = \frac{1}{(2 (\Delta p_0)^2)^{1/4}} e^{- i(p-
p_0)x_0/\hbar} e^{- (p - p_0)^2/4 (\Delta p_0)^2}, \label{mo wa}
\end{equation}
where the variances of the position and momentum are related by
the minimum uncertainty
\begin{equation}
\Delta x_0 \Delta p_0 = \frac{\hbar}{2}. \label{un re}
\end{equation}
The Gaussian wave packet (\ref{ga wa}) has the expectation values
and the variances of the position and momentum
\begin{eqnarray}
\langle \psi(x, 0) \vert \hat{x} \vert \psi(x, 0) \rangle &=& x_0,
\quad \langle \psi(x, 0) \vert \hat{p} \vert \psi(x, 0) \rangle =
p_0, \nonumber\\
\langle \psi(x, 0) \vert (\hat{x} - \langle \hat{x} \rangle)^2
\vert \psi(x, 0) \rangle &=& (\Delta x_0)^2, \quad \langle \psi(x,
0) \vert  (\hat{p} - \langle \hat{p} \rangle)^2 \vert \psi(x, 0)
\rangle = (\Delta p_0)^2.
\end{eqnarray}

However, as far as the uncertainty principle is satisfied, one can
have four parameters to fix the profile of a general Gaussian wave
packet at the initial time: two expectation values and two
variances
\begin{eqnarray}
x_0 &=& \langle \hat{x} \rangle_0, \quad p_0 = \langle \hat{p}
\rangle_0, \nonumber\\
(\Delta x_0)^2 &=& \langle (\hat{x} - x_0)^2 \rangle_0, \quad
(\Delta p_0)^2 = \langle (\hat{p} - p_0)^2 \rangle_0. \label{va}
\end{eqnarray}
A simple way to make the momentum variance independent of the
position variance without changing the probability distribution in
the position space would introduce a phase factor \cite{howard}
\begin{equation}
\tilde{\psi} (x, t = 0) = e^{i c x^2/\hbar} \psi(x, t= 0).
\end{equation}
There is no new parameter in the wave function of the
Schr\"{o}dinger equation except for the four parameters in the
initial wave packet, since the evolution of the wave packet is
entirely determined by
\begin{equation}
\Psi(x, t) = e^{- i \hat{H} t/ \hbar} \psi(x, 0) = \int
\frac{dp}{(2 \pi \hbar)^{1/2}} e^{i px/ \hbar} \phi(p, t = 0).
\label{sch wa}
\end{equation}

On the other hand, in the Heisenberg picture, an operator evolves
according to the Heisenberg equation
\begin{equation}
i \hbar \frac{d \hat{A}}{d t}  = i \hbar \frac{\partial \hat{A}
}{\partial t} + [ \hat{A}, \hat{H}].
\end{equation}
For a free mass, the Heisenberg equations become
\begin{eqnarray}
\frac{d \langle \hat{x} \rangle}{dt} &=& \frac{\langle \hat{p}
\rangle}{m}, \quad \frac{d \langle \hat{p} \rangle}{dt} = 0,
\nonumber\\
\frac{d^2 \langle \hat{x}^2 \rangle}{dt^2} &=& \frac{\langle
\hat{p}^2 \rangle}{m^2}, \quad \frac{d \langle \hat{p}^2
\rangle}{dt} = 0.
\end{eqnarray}
Then the expectation values of the position and momentum at any
time  can be found as \cite{yuen,styer}
\begin{eqnarray}
\langle \hat{x} \rangle_t &=& \langle \hat{x} \rangle_0 + \langle
\hat{p} \rangle_0 \frac{t}{m}, \nonumber\\
\langle \hat{p} \rangle_t &=& \langle \hat{p} \rangle_0, \label{fr
ex so}
\end{eqnarray}
and the variances as
\begin{eqnarray}
(\Delta x_t)^2 &=& (\Delta x_0)^2 + \{\langle (\hat{x} - \langle x
\rangle_0)(\hat{p} - \langle \hat{p} \rangle_0) + (\hat{p} -
\langle \hat{p} \rangle_0) (\hat{x} - \langle x
\rangle_0)\rangle_0\}
\frac{t}{m} + (\Delta p_0)^2\frac{t^2}{m^2}, \nonumber\\
(\Delta p_t)^2 &=& (\Delta p_0)^2. \label{fr va}
\end{eqnarray}
One can see from (\ref{fr ex so}) and (\ref{fr va}) that there are
four independent initial data: $\langle \hat{x} \rangle_0, \langle
\hat{p} \rangle_0, \Delta x_0$ and $\Delta p_0$. A similar
argument holds for the oscillator \cite{styer}. The linearly
time-dependent term in (\ref{fr va}) was introduced earlier in
\cite{gottfried} and was discussed later in the context of
coherent states \cite{yuen,wodkiewicz}. It should be noted that
the linearly time-dependent term is related with the correlation
between the noncommuting operators  $\hat{x}$ and $\hat{p}$
\cite{bohm}.

\section{Invariant Operator Method}

An immediate question may be asked which wave packet has the four
parameters (\ref{va}) to fit the initial profile. Instead of
directly looking for such a wave packet, we go around by solving
the time-dependent Schr\"{o}dinger equation
\begin{equation}
i \hbar \frac{\partial}{ \partial t} \Psi (x, t) = \hat{H} (t)
\Psi, \label{td sc}
\end{equation}
for the free mass or the oscillator
\begin{eqnarray}
\hat{H}_f = \frac{\hat{p}^2}{2m}, \quad \hat{H}_o =
\frac{\hat{p}^2}{2m} + \frac{m \omega^2}{2} \hat{x}^2. \label{os
ha}
\end{eqnarray}
Though it would be straightforward to solve (\ref{td sc}) for the
free mass and the oscillator, we shall use the invariant operator
method by Lewis and Riesenfeld \cite{lewis}. In the invariant
operator method one first solves the quantum Liouville-von Neumann
equation
\begin{equation}
i \hbar \frac{\partial \hat{I}}{\partial t}  + [\hat{I}, \hat{H}]
= 0, \label{ln eq}
\end{equation}
and then finds the eigenfunction
\begin{equation}
\hat{I}(t) \psi_{\lambda}(x, t) = \lambda \psi_{\lambda}(x, t).
\label{ei eq}
\end{equation}
Then the wave function of the Schr\"{o}dinger equation is given by
\begin{equation}
\Psi_{\lambda} (x, t) = e^{ i \int \langle \psi_{\lambda} (t)
\vert i
\partial / \partial t - \hat{H}/{\hbar} \vert \psi_{\lambda} (t)
\rangle dt} \psi_{\lambda} (x, t), \label{in wa}
\end{equation}
Though the invariant operator method was originally introduced to
study time-dependent quantum systems, it can be used even for
time-independent systems such as (\ref{os ha}).

It is known that there are a pair of invariant operators of the
form \cite{dod,kim1,kim-lee,nieto,kim-page,kim2,kim3}
\begin{eqnarray}
\hat{a} (t) = \frac{i}{\sqrt{\hbar}} (u^* (t) \hat{p} - m
\dot{u}^* (t) \hat{x}), \quad \hat{a}^{\dagger} (t) = -
\frac{i}{\sqrt{\hbar}} (u (t) \hat{p} - m \dot{u} (t) \hat{x}),
\label{c-a}
\end{eqnarray}
where $u$ is a complex solution of the motion either for the free
mass
\begin{equation}
\ddot{u} = 0, \label{fr eq}
\end{equation}
or for the oscillator
\begin{equation}
\ddot{u} + \omega^2 u = 0. \label{os eq}
\end{equation}
When the wronskian condition is imposed
\begin{equation}
m (u \dot{u}^* - \dot{u} u^*) = i, \label{wr co}
\end{equation}
the invariant operators (\ref{c-a}) become the time-dependent
annihilation (lowering) and creation (raising) operators that
satisfy the equal time commutation relation
\begin{equation}
[ \hat{a} (t), \hat{a}^{\dagger} (t) ] = 1.
\end{equation}
For instance, the complex solution to the oscillator
\begin{equation}
u_0 (t) = \frac{e^{- i \omega t}}{\sqrt{2 m \omega}}, \label{os
so}
\end{equation}
satisfies (\ref{wr co}) and leads to the invariant operators
\begin{equation}
\hat{a} (t) = e^{i \omega t} \hat{a}, \quad \hat{a}^{\dagger} (t)
= e^{-i \omega t} \hat{a}^{\dagger},
\end{equation}
where $\hat{a}$ and $\hat{a}^{\dagger}$ are the standard
annihilation and creation operators
\begin{eqnarray}
\hat{a} = \frac{1}{\sqrt{2 \hbar}} \Bigl(\sqrt{m \omega} \hat{x} +
\frac{i}{\sqrt{m \omega}} \hat{p} \Bigr), \quad \hat{a}^{\dagger}
= \frac{1}{\sqrt{2 \hbar}} \Bigl(\sqrt{m \omega} \hat{x} -
\frac{i}{\sqrt{m \omega}} \hat{p} \Bigr).
\end{eqnarray}
Note that the invariant operators have the opposite sign in the
frequency from that of the Heisenberg operators
\begin{equation}
\hat{a}_H (t) = e^{-i \omega t} \hat{a}, \quad \hat{a}^{\dagger}_H
(t) = e^{i \omega t} \hat{a}^{\dagger}.
\end{equation}

From now on we shall find a more general solution to (\ref{fr eq})
and (\ref{os eq}). The general solution can be found in the form
\begin{equation}
u_{r} (t) = (\cosh r) u_0(t) + (e^{-i \vartheta} \sinh r) u_0^*
(t), \quad (r \geq 0, ~ 2 \pi > \vartheta \geq 0) \label{sq so}
\end{equation}
where $u_0$ is a preferred solution satisfying (\ref{wr co}) for
the free mass
\begin{equation}
u_0(t) = \frac{1}{\sqrt{2}} \Bigl(1 - i\frac{t}{m} \Bigr),
\label{fr so}
\end{equation}
and (\ref{os so}) for the oscillator. Note that there are two
arbitrary real parameters $r$ and $\vartheta$ in the general
solution (\ref{sq so}) which are required for a second order
differential equation. The general solution ({\ref{sq so}) leads
to a Bogoliubov transformation between the operators $\hat{a}_r$
and $\hat{a}_{r}^{\dagger}$ obtained by putting $u_r$ and those
operators $\hat{a}_0$ and $\hat{a}_{0}^{\dagger}$ obtained by
putting $u_0$ in (\ref{c-a}):
\begin{eqnarray}
\hat{a}_{r} (t)&=& (\cosh r) \hat{a}_0 (t) - (e^{i \vartheta}
\sinh r)
\hat{a}_0^{\dagger} (t), \nonumber\\
\hat{a}_{r}^{\dagger} (t) &=& (\cosh r) \hat{a}_0^{\dagger} (t) -
(e^{- i \vartheta} \sinh r) \hat{a}_0 (t).
\end{eqnarray}
Here $r$ is a squeeze parameter and $\vartheta$ is a squeeze
angle. In fact, in terms of the squeeze operator
\begin{equation}
\hat{S} (z, t) = e^{(z \hat{a}_0^2 (t) - z^* \hat{a}_0^{\dagger
2})/2}, \quad z = e^{i (\pi -\vartheta)} r,
\end{equation}
$\hat{a}_r$ can be written as a unitary transformation
\cite{stoler}
\begin{equation}
\hat{a}_r (t) = \hat{S} (z, t) \hat{a}_0 (t) \hat{S}^{\dagger} (z,
t). \label{un tr}
\end{equation}
Hence any wave function obtained by using $\hat{a}_r$ and
$\hat{a}_r^{\dagger}$ is the squeezed state of the corresponding
wave function by $\hat{a}$ and $\hat{a}^{\dagger}$. For instance,
the number state $\psi_{nr}$ of $\hat{N}_r = \hat{a}_r^{\dagger}
\hat{a}_r$ is the squeezed number state of $\psi_{n0}$ of
$\hat{N}_0 = \hat{a}_0^{\dagger} \hat{a}_0$, that is, $\psi_{nr} =
\hat{S} \psi_{n0}$.

\section{Coherent-Squeezed States}

To find the general Gaussian wave packet, we introduce a coherent
state
\begin{equation}
\hat{a}_r (t) \psi_{\alpha r} (x, t) = \alpha \psi_{\alpha r} (x,
t), \label{co st}
\end{equation}
for a complex number $\alpha$. It follows from (\ref{un tr}) that
$\psi_{\alpha r}$ is the squeezed state of the coherent state
$\psi_{\alpha 0}$ defined by $\hat{a}_0$. In fact, the state
(\ref{co st}) is the coherent-squeezed state (ideal squeezed
state) \cite{caves}
\begin{equation}
\psi_{\alpha r} (x, t) = \hat{D} (\alpha, t) \hat{S} (z,t)
\psi_{00} (x, t),
\end{equation}
where $\hat{D} = e^{\alpha \hat{a}^{\dagger}_r - \alpha^*
\hat{a}_r}$ is the displacement operator and $\psi_{00}$ is the
ground state of $\hat{a}^{\dagger}_0 \hat{a}_0$. Note that the
coherent-squeezed state (\ref{co st}) has one complex constant
$\alpha$, two real constants $r$ and $\vartheta$, which will be
related with four parameters in (\ref{va}). From the position and
momentum operators
\begin{eqnarray}
\hat{x} &=& \sqrt{\hbar} (u_r (t) \hat{a}_r (t) + u_r^* (t)
\hat{a}_r^{\dagger}(t)), \nonumber\\
\hat{p} &=& \sqrt{\hbar} m (\dot{u}_r (t) \hat{a}_r (t) +
\dot{u}_r^* (t) \hat{a}_r^{\dagger}(t)),
\end{eqnarray}
their expectation values with respect to the coherent state
(\ref{co st}) yield
\begin{eqnarray}
\langle \psi_{\alpha r} (t) \vert \hat{x} \vert \psi_{\alpha r}
(t) \rangle
&=& \sqrt{\hbar} ( u_r(t) \alpha + u_r^*(t) \alpha^* ), \nonumber\\
\langle \psi_{\alpha r} (t) \vert \hat{p} \vert \psi_{\alpha r}
(t) \rangle &=& \sqrt{\hbar} m (\dot{u}_r(t) \alpha +
\dot{u}_r^*(t) \alpha^*). \label{p-m ex}
\end{eqnarray}
We shall choose $\alpha$ for the free mass
\begin{equation}
\alpha = \frac{1}{\sqrt{2 \hbar}} \{ (\cosh r  - e^{i \vartheta}
\sinh r)x_0 + i (\cosh r  + e^{i \vartheta} \sinh r) p_0 \},
\label{a1}
\end{equation}
and for the oscillator
\begin{equation}
\alpha = \frac{1}{\sqrt{2 \hbar}} \{ \sqrt{m \omega}(\cosh r  -
e^{i \vartheta} \sinh r)x_0 + \frac{i}{\sqrt{m \omega}} (\cosh r +
e^{i \vartheta} \sinh r) p_0 \}. \label{a2}
\end{equation}
Then the expectation values follow the classical trajectory
\begin{equation}
\langle \psi_{\alpha r} (t) \vert \hat{x} \vert \psi_{\alpha r}
(t) \rangle = x_c (t), \quad \langle \psi_{\alpha r} (t) \vert
\hat{p} \vert \psi_{\alpha r} (t) \rangle = p_c (t), \label{p-m
ex2}
\end{equation}
where the classical position $x_c$ for the free mass or the
oscillator is, respectively,
\begin{eqnarray}
x_c (t) &=& x_0 + p_0\frac{t}{m}, \nonumber\\ x_c (t) &=& x_0 \cos
(\omega t) + \frac{p_0}{m \omega} \sin (\omega t). \label{cl tr}
\end{eqnarray}
The classical momentum is simply given by $p_c = m \dot{x}_c$.
Note that the classical position and momentum have the initial
values $x_c (0) = x_0$ and $p_c (0) = p_0$ as expected.

Finally, we obtain the wave packet for the coherent state (\ref{co
st})
\begin{eqnarray}
\Psi_{\alpha r} (x, t) = \Bigl(\frac{1}{ \sqrt{2\pi \hbar} u_r^*}
\Bigr)^{1/2} e^{- i S_c / \hbar} e^{i p_c x/\hbar} e^{i m
\dot{u}_r^* (x - x_c)^2 / 2 \hbar u_r^* }, \label{wa pa}
\end{eqnarray}
where $S_c$ is the classical action
\begin{eqnarray}
S_c (t) = \int^t L_c (t') dt'.
\end{eqnarray}
The classical Lagrangian $L_c$ for the free mass is
\begin{equation}
L_c (t) = \frac{p_c^2(t)}{2m} = \frac{p_0^2}{2m},
\end{equation}
and for the oscillator
\begin{equation}
L_c (t) = \frac{p_c^2 (t)}{2m} - \frac{m \omega^2}{2} x^2_c (t) =
\Bigl(\frac{p_0^2}{2m} - \frac{m \omega^2 x_0^2}{2} \Bigr) \cos(2
\omega t) - \omega x_0 p_0 \sin (2 \omega t) .
\end{equation}
As $e^{- i S_c/ \hbar}$ is a pure phase factor, the centroid of
the Gaussian wave packet (\ref{wa pa}) follows the same classical
trajectory as expected of a coherent state:
\begin{equation}
\langle \Psi_{\alpha r} (t) \vert \hat{x} \vert \Psi_{\alpha r}
(t) \rangle = x_c (t), \quad \langle \Psi_{\alpha r} (t) \vert
\hat{p} \vert \Psi_{\alpha r} (t) \rangle = p_c (t).
\end{equation}
Thus we have obtained the travelling general wave packet (\ref{wa
pa}) whose centroid is the classical trajectory (\ref{cl tr}).

\section{Squeeze Parameters}

The free mass has the time-dependent variance of the position with
respect to the Gaussian wave packet (\ref{wa pa})
\begin{eqnarray}
(\Delta x_t)^2 = \hbar u_r^* (t) u_r (t)  = \frac{\hbar}{2}
\Bigl\{ e^{2r} \Bigl(\cos \frac{\vartheta}{2} + \frac{t}{m} \sin
\frac{\vartheta}{2} \Bigr)^2 + e^{-2r} \Bigl(\sin
\frac{\vartheta}{2} - \frac{t}{m} \cos \frac{\vartheta}{2}
\Bigr)^2 \Bigr\},
\end{eqnarray}
and the constant variance of the momentum
\begin{eqnarray}
(\Delta p_t)^2= \hbar m^2 \dot{u}_r^* (t) \dot{u}_r (t) =
\frac{\hbar}{2} \Bigl( e^{2r} \sin^2 \frac{\vartheta}{2} + e^{-2r}
\cos^2 \frac{\vartheta}{2} \Bigr).
\end{eqnarray}
Similarly, the oscillator has the time-dependent variances of the
position and momentum
\begin{eqnarray}
(\Delta x_t)^2 &=& \frac{\hbar}{2m \omega} \{ \cosh (2r) + \sinh
(2r) \cos (2 \omega t - \vartheta)\},
\nonumber\\
 (\Delta p_t)^2 &=& \frac{\hbar m \omega}{2} \{ \cosh (2r) - \sinh
(2r) \cos (2 \omega t - \vartheta)\}.
\end{eqnarray}
However, the correlation between the position and momentum is not
an independent quantity but is determined either by $r$ and
$\vartheta$ or $\Delta x_t$ or $\Delta p_t$ since the parameters
$r$ and $\vartheta$ are determined by the two variances or vice
versa. The position-momentum correlation
\begin{eqnarray}
\Delta (xp)_t = \frac{1}{2} \langle (\hat{p} - p_c) \hat{x} - x_c)
+ (\hat{x} - x_c) (\hat{p} - p_c) \rangle = \frac{\hbar m}{2}
(\dot{u}_r^*(t) u_r (t) + u_r^* (t) \dot{u}_r (t))
\end{eqnarray}
is for the free mass
\begin{equation}
\Delta (xp)_t = \frac{\hbar}{2} \Bigl\{\sinh (2r) \sin \vartheta +
(\cosh (2r) - \sinh (2r) \cos \vartheta) \frac{t}{m} \Bigr\},
\end{equation}
and for the oscillator
\begin{equation}
\Delta (xp)_t = - \frac{\hbar}{2} \sinh (2r) \sin(2 \omega t -
\vartheta).
\end{equation}

Now we determine the squeeze parameter $r$ and squeeze angle
$\vartheta$ in terms of the initial variances at $t = 0$. The
parameters $r$ and $\vartheta$ for the free mass are given by
\begin{eqnarray}
\cosh (2r) &=& \frac{1}{\hbar} \{(\Delta x_0)^2 + (\Delta
p_0)^2\},
\nonumber\\
\cos \vartheta \sinh (2r) &=& \frac{1}{\hbar} \{(\Delta x_0)^2 -
(\Delta p_0)^2 \},  \label{sq pa}
\end{eqnarray}
and for the oscillator are given by
\begin{eqnarray}
\cosh (2r) = \frac{1}{\hbar} \Bigl\{ m \omega (\Delta x_0)^2 +
\frac{1}{m \omega}(\Delta p_0)^2 \Bigr\},
\nonumber\\
\cos \vartheta \sinh (2r) = \frac{1}{\hbar} \Bigl\{m \omega
(\Delta x_0)^2 - \frac{1}{m \omega}(\Delta p_0)^2 \Bigr\}.
\label{sq pa2}
\end{eqnarray}
In both cases of the free mass and the oscillator, an important
dimensionless parameter that denotes the amount of deviation from
the minimum uncertainty or the position-momentum correlation at
the initial time can be introduced
\begin{eqnarray}
\sin \vartheta \sinh (2r) = \frac{2}{\hbar} \Delta (xp)_0 = \pm
\sqrt{(2\Delta x_0\Delta p_0/\hbar)^2 -1}  \equiv \pm \delta .
\label{sq pa3}
\end{eqnarray}
It should be noted that the position-momentum correlation is
determined by the variances of the position and momentum as
expected. Here and in this paper it is assumed that the variances
satisfy the uncertainty principle
\begin{equation}
\Delta x_0 \Delta p_0 \geq \frac{\hbar}{2}.
\end{equation}

Using (\ref{sq pa}) or (\ref{sq pa2}), it is easy to show that the
initial general Gaussian wave packet for both the free mass and
the oscillator takes the form
\begin{eqnarray}
\Psi_{\alpha r} (x, 0) = \Bigl(\frac{1}{ \sqrt{2\pi \hbar}
u_r^*(0)} \Bigr)^{1/2} e^{i p_0 x/\hbar} e^{- (1 \mp i \delta ) (x
- x_0)^2 / 4 (\Delta x_0)^2}. \label{in pa}
\end{eqnarray}
For the minimum uncertainty, $(\Delta x_0 \Delta p_0 = \hbar/2$
and so $\delta = 0)$, we recover the wave packet (\ref{sch wa})
with the initial distribution (\ref{ga wa}) and (\ref{mo wa}).
However, the minimum uncertainty does not always imply $r = 0$,
since one may choose $\vartheta =0$ and $e^{r} = \sqrt{2/\hbar}
\Delta x_0$ for the free mass and $e^{r} = \sqrt{2 m \omega/\hbar}
\Delta x_0$ for the oscillator, or $\vartheta = \pi$ and $e^{r} =
\sqrt{2/\hbar} \Delta p_0$ for the free mass and $e^{r} =
\sqrt{2/\hbar m \omega} \Delta p_0$ for the oscillator. The
minimum energy will select $r = 0$ and $\vartheta = 0$ for both
the free mass and oscillator. Using (\ref{wr co}), we write the
exponent of (\ref{wa pa}) as
\begin{equation}
\frac{m \dot{u}_r^*}{\hbar u_r^*} = \frac{i + m (\dot{u}_r^* u_r +
u_r^* \dot{u}_r)}{2 \hbar u_r^* u_r} = \frac{i + 2 \Delta (xp)_t/
\hbar}{2 (\Delta x_t)^2},
\end{equation}
and, finally, obtain the travelling wave packet in the form
\begin{eqnarray}
\Psi_{\alpha r} (x, t) = \Bigl(\frac{1}{ \sqrt{2\pi \hbar}
u_r^*(t)} \Bigr)^{1/2} e^{- i S_c(t)/ \hbar} e^{i p_c (t) x/\hbar}
e^{- (1 - 2 i \Delta (xp)_t / \hbar) (x - x_c(t))^2 / 4 (\Delta
x_t)^2}. \label{va pa}
\end{eqnarray}

From (\ref{sq pa}), (\ref{sq pa2}) and (\ref{sq pa3}), the
variances at any time can be found for the free mass
\begin{eqnarray}
(\Delta x_t)^2 = (\Delta x_0)^2 \pm \hbar \delta \frac{t}{m} +
(\Delta p_0)^2 \frac{t^2}{m^2}, \label{td va}
\end{eqnarray}
and for the oscillator
\begin{eqnarray}
(\Delta x_t)^2 = (\Delta x_0)^2 \cos^2 (\omega t) \pm \frac{\hbar
\delta }{2m \omega} \sin(2 \omega t) + \frac{(\Delta p_0)^2}{(m
\omega)^2} \sin^2 (\omega t). \label{td va2}
\end{eqnarray}
The linearly time-dependent term in (\ref{td va}) or (\ref{td
va2}) was interpreted due to the probability current of the
initial wave packet \cite{gottfried,wodkiewicz}. Later Yuen solved
the Heisenberg equation for the position variance using an initial
coherent-squeezed state to get such a linearly time-dependent
term, in particular, with the negative sign \cite{yuen}. Here we
have found the travelling general wave packet from the initial one
and thereby have obtained the variances. It is not hard to show
that (\ref{sq pa3}) comes from the correlation between $\hat{x}$
and $\hat{p}$ at the initial time.

There are two interesting cases of minimal- and
maximal-uncertainty states. First, the variance of the position in
the minimal-uncertainty $(\delta = 0)$ states monotonically
increases for the free mass
\begin{eqnarray}
(\Delta x_t)^2 = (\Delta x_0)^2 + (\Delta p_0)^2 \frac{t^2}{m^2},
\end{eqnarray}
but oscillates for the oscillator
\begin{eqnarray}
(\Delta x_t)^2 = (\Delta x_0)^2 \cos^2 (\omega t) + \frac{(\Delta
p_0)^2}{(m \omega)^2} \sin^2 (\omega t).
\end{eqnarray}
The spreading of the wave packet for the free particle is
inevitable for minimal-uncertainty states. For
nonminimal-uncertainty $(\delta > 0)$ state, however, there is one
interesting case for the negative sign of (\ref{td va}). Though
the late time evolution of variance is dominated by the last term,
there is an interim when the variance first decreases until
\begin{eqnarray}
\tau = \frac{\hbar m \delta}{2 \Delta p_0^2}, \label{tau}
\end{eqnarray}
and then increases to the initial value $\Delta x_0$ at $t = 2
\tau$. The minimum value of the variance is
\begin{equation}
\Delta x_{\tau}^2 = \frac{3 \Delta x_0^2}{4} + \frac{\hbar^2}{4
\Delta p_0^2}.
\end{equation}
Second, the maximal-uncertainty $(\delta \gg 1)$ states, have
approximately the position variance for the free mass
\begin{eqnarray}
\Delta x_t \approx \Bigl| \Delta x_0 \pm  \frac{\hbar \delta}{2
\Delta x_0} \frac{t}{m} \Bigr|, \label{max1}
\end{eqnarray}
and for the oscillator
\begin{eqnarray}
\Delta x_t \approx \Bigl| \Delta x_0 \cos (\omega t) \pm
\frac{\hbar \delta}{2 m \omega \Delta x_0} \sin (\omega t) \Bigr|.
\label{max2}
\end{eqnarray}
For maximal-uncertainty states one has $\Delta x_{\tau} \approx
\sqrt{3} \Delta x_0/2$. It is thus possible, in principle, to
measure the position the second time without much disturbance from
the first measurement \cite{yuen}.

\section{Conclusion}

Gaussian wave packets have many applications, for instance, in
quantum optics or atomic physics or quantum measurement theory. In
particular, the quantum motion of the measuring apparatus of a
free mass or an oscillator will be monitored to detect the
gravitational waves. The general Gaussian wave packet can have
four parameters to be fixed: the expectation values, $x_0$ and
$p_0$, and the variances, $\Delta x_0$ and $\Delta p_0$ of the
position and momentum. The only constraint on the variances is the
uncertainty principle, $\Delta x_0 \Delta p_0 \geq \hbar/2$.

In this paper we have found a coherent-squeezed state
representation of the travelling general Gaussian wave packet. The
squeeze parameter and angle for a squeezed state are determined by
the initial variances of the position and momentum, which are
given by (\ref{sq pa}) for the free mass and by (\ref{sq pa2}) for
the oscillator. Further, the complex parameter for the coherent
state is related with the centroid of the Gaussian wave packet and
is determined by (\ref{a1}) for the free mass and by (\ref{a2})
for the oscillator. Thus we have been able to obtain the
travelling general Gaussian wave packet (\ref{va pa}) whose
centroid moves along the classical trajectory with $x_c (t)$ and
$p_c (t)$. The minimal-uncertainty states keep the well-known form
whereas all the other nonminimal-uncertainty states have an
additional position-squared phase proportional to a dimensionless
measure of deviation from the minimum uncertainty or the initial
position-momentum correlation. The effect of this phase is to
correlate the initial position and momentum and modify the time
evolution of the variances as given by (\ref{td va}) for the free
mass and (\ref{td va2}) for the oscillator.

An important consequence of the travelling general Gaussian wave
packet (\ref{va pa}) is that all non-minimal uncertainty states
will have an additional contribution to the variances of the
position, the linearly time-dependent term in (\ref{td va}) for
the free mass and in (\ref{td va2}) for the oscillator. There
exists a specific initial Gaussian profile (\ref{va pa}) with the
negative sign for the free mass that will contribute negatively to
the position variance and result in a minimum variance at a
certain moment (\ref{tau}). Yuen proposed such a state, the
so-called contractive state, and used it to test the standard
limit for monitoring the free mass system \cite{yuen}. However,
the state was defined only at the initial time and the evolution
of variances was found in the Heisenberg picture. In our
coherent-squeezed state representation, the wave packet is the
travelling general Gaussian wave packet from an initial general
Gaussian one. It would be interesting to investigate how to
prepare such a state with minimal variance of the position and how
the system evolves during measurements. The travelling general
wave packets are also expected to have useful applications in
studying quantum revivals and collapses.

\acknowledgements

This work was supported by the Korea Research Foundation under
grant No. KRF-2004-041-C00480.

\end{document}